\documentclass[review]{elsarticle}

\usepackage{hyperref}

\journal{Journal of \LaTeX\ Templates}

\usepackage{xcolor}
\usepackage{graphicx}
\usepackage{bm}
\usepackage{xcolor}
\usepackage{makecell}

\usepackage{subfigure}
\usepackage{amsmath}
\usepackage{multirow}
\usepackage{amsfonts}       
\usepackage{diagbox}
\usepackage{booktabs}
\usepackage{algorithm}
\usepackage{times}

\begin{document}

\begin{frontmatter}


\title{Blind Quality Assessment for Image Superresolution Using Deep Two-Stream Convolutional Networks}

\author{Wei Zhou, Qiuping Jiang, Yuwang Wang, Zhibo Chen, Weiping Li}






\begin{abstract}
  Numerous image superresolution (SR) algorithms have been proposed for reconstructing high-resolution (HR) images from input images with lower spatial resolutions. However, effectively evaluating the perceptual quality of SR images remains a challenging research problem. In this paper, we propose a no-reference/blind deep neural network-based SR image quality assessor (DeepSRQ). To learn more discriminative feature representations of various distorted SR images, the proposed DeepSRQ is a two-stream convolutional network including two subcomponents for distorted structure and texture SR images. Different from traditional image distortions, the artifacts of SR images cause both image structure and texture quality degradation. Therefore, we choose the two-stream scheme that captures different properties of SR inputs instead of directly learning features from one image stream. Considering the human visual system (HVS) characteristics, the structure stream focuses on extracting features in structural degradations, while the texture stream focuses on the change in textural distributions. In addition, to augment the training data and ensure the category balance, we propose a stride-based adaptive cropping approach for further improvement. Experimental results on three publicly available SR image quality databases demonstrate the effectiveness and generalization ability of our proposed DeepSRQ method compared with state-of-the-art image quality assessment algorithms.
\end{abstract}

\begin{keyword}
Image superresolution \sep blind image quality assessment \sep two-stream convolutional networks \sep stride-based adaptive cropping \sep human vision
\end{keyword}

\end{frontmatter}

\section{Introduction}\label{sec:introduction}
Image superresolution (SR) aims at constructing a high-resolution (HR) image with fine details using one or several low-resolution (LR) images as inputs \cite{tsai1984multiframe}, which is desirable in various practical scenarios, such as medical imaging, video surveillance, and high-definition television (HDTV) \cite{park2003super}. Through SR technologies, people can better view LR images on HR displays. During the past few decades, many generic single image superresolution algorithms have been proposed \cite{yan2019deep}. However, much less has been done to fairly evaluate the perceptual quality of superresolved images (SRIs) and the performance of SR algorithms \cite{baker2002limits}.

For image superresolution quality assessment, in the literature, small-scale subjective experiments are usually used for evaluation. Specifically, subjects are asked to rate the visual quality of SR images generated by different SR algorithms. The rating provided for each SR image under examination is termed the opinion score. The mean of these ratings, i.e., the mean opinion score (MOS), is calculated as the ground-truth image quality measurement, which is a common practice in quality assessment. Thus, although some viewers may have different feelings, the MOS score is a statistical concept. Subjective tests were performed in \cite{yang2014single} to explore the visually subjective quality of SR images. Such subjective testing is reliable for providing a fair evaluation of SR image quality but is expensive, labor intensive, and time consuming. More importantly, subjective tests cannot be integrated into the automatic design process of perception-driven SR algorithms. Therefore, it is desirable to develop objective IQA methods for automatically predicting the subjective visual quality of SR images.

When the original distortion-free image is available, full-reference image quality assessment (FR-IQA) can be carried out by comparing the distorted image with the reference image. Two classic FR-IQA metrics, namely, the peak signal-to-noise-ratio (PSNR) and the structural similarity index (SSIM) \cite{wang2004image}, are usually employed to evaluate the visual perceptual quality of reconstructed SR images. However, they do not match well with a subjective evaluation. Three publicly available SR image quality databases considering several typical SR algorithms have been built \cite{ma2017learning,wang2017perceptual,zhou2019visual}, upon which state-of-the-art IQA metrics are tested. The results demonstrate that it is difficult for the existing IQA metrics to effectively predict the perceptual quality of SR images.
In addition, reference images seldom exist in most situations, and these FR-IQA metrics require the resolution of the distorted image to be the same as that of the original image. As a result, a no-reference/blind image quality assessment (NR-IQA) metric specifically designed for image superresolution, which evaluates perceptual quality with no access to reference LR or HR images, is directly applicable and highly demanded.


\begin{figure}[t]
  \centerline{\includegraphics[width=1.0\linewidth]{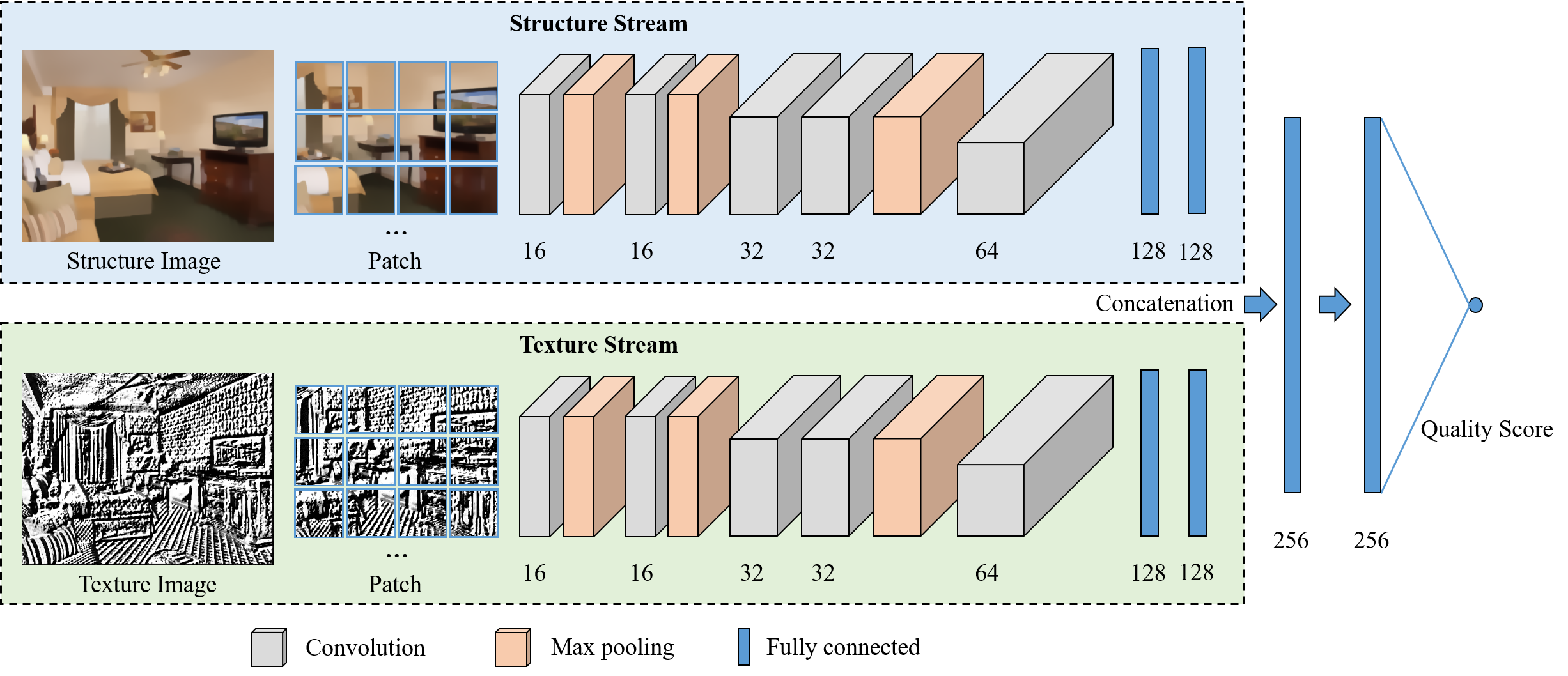}}
  \caption{Overview of the network architecture of our proposed model for learning the perceptual quality of image superresolution. In DeepSRQ, representative features are extracted from distorted structure and texture SR images by two subnetworks. The extracted features are then used to estimate the quality scores by late fusion regression layers.}
  \centering
\label{fig:fig2}
\end{figure}

In recent years, deep learning-related methods have been extensively used and have achieved great achievements for a variety of image processing and computer vision problems, such as image recognition \cite{he2016deep}, video quality assessment \cite{zhou2018stereoscopic}, and social image understanding \cite{li2018deep}. In this paper, we exploit deep convolutional neural networks (DCNNs) to address the no-reference image superresolution quality assessment problem. Specifically, we propose a deep neural network-based SR image quality assessor (DeepSRQ). Inspired by \cite{zhou2019visual}, the proposed model shown in Figure \ref{fig:fig2} contains two subcomponents accounting for structure and texture SR images. To the best of our knowledge, for objective SR IQA models, this is the first study of applying the two-stream network architecture to learn the perceptual SR image quality. The human visual system (HVS) is not only sensitive to structural information but also considers textural details \cite{aujol2006structure}. Moreover, the distortions of SR images are different from conventional artifacts, which degenerate both image structure and texture \cite{zhou2019visual}. Thus, we first extract structure and texture images from distorted SR images. Second, since existing image superresolution quality databases (i.e., the superresolution (SR) quality database \cite{ma2017learning}, superresolution reconstructed image database (SRID) \cite{wang2017perceptual}) and quality assessment database for SRIs (QADS)) are relatively small-scale, and we opt to utilize a stride-based adaptive cropping approach to augment the training data and ensure the category balance. This approach also considers the local visual information of the whole distorted structure and texture SR images. Third, we take the distorted structure and texture SR patches as inputs and train the two-stream convolutional network to extract the discriminative feature representations. Fourth, different from the original classification-based DCNN \cite{krizhevsky2012imagenet}, we aim to map the feature representations to estimated scores, which adopts a fully connected layer instead of a softmax layer. Specifically, two fully connected layers are followed by each substream. Then, we use concatenation to further obtain one quality score for each SR image patch. Finally, we average these estimated scores as the perceptual quality for the entire SR image.

In addition, the training process of DeepSRQ produces both low-level visual information and high-level semantic features. Our experimental results show that the synthetically learned features are more effective than both the handcrafted low-level features and high-level semantic features extracted from pretrained DCNN models. Each substream is verified in our experiments. We also conducted an ablation study to demonstrate that the proposed stride-based adaptive cropping approach indeed plays a critical role in DeepSRQ.

The contributions of this work are summarized as follows:

\begin{itemize}
\item Since the distortions of SR images cause both image structure and texture quality degradation, we propose a two-stream deep convolutional network for the blind quality estimation of superresolution images. The proposed network extracts the discriminative features from various distorted structure and texture SR images, where each subnetwork adapts and differs from the classification-based architecture.
\item To ensure the category balance, we propose a stride-based adaptive cropping approach for augmenting the training data. We show the effectiveness of the proposed adaptive cropping method for further improving the performance of the whole framework.
\item We conduct extensive experiments on various databases demonstrating the effectiveness of the proposed network and the corresponding adopted techniques. In addition, the synthetic features learned from the proposed two-stream network are more effective than traditional features.
\end{itemize}

The remainder of this paper is organized as follows. In Section II, we introduce the proposed deep neural network-based SR image quality assessor (DeepSRQ) for no-reference/blind superresolution image quality prediction in detail. Section III presents the experimental results and analysis. In Section IV, we conclude the paper and discuss future research directions.

\section{Proposed DeepSRQ}
The network architecture of the proposed blind quality assessor for learning the perceptual quality of image superresolution is shown in Figure \ref{fig:fig2}. We design our proposed DeepSRQ network with a two-stream architecture, which takes distorted structure and texture SR images as inputs through two subnetworks, i.e., structure stream and texture stream. Each subnetwork adapts and differs from AlexNet \cite{krizhevsky2012imagenet}. It replaces the last softmax layer with a fully connected layer for the regression task. We thus adopt MSE loss rather than cross-entropy loss. The whole network design, such as kernel number and network parameters, is also different. Since existing image superresolution quality databases are relatively small-scale, we opt to augment the training set by cropping the distorted structure and texture SR images into multiple patches. In addition, to ensure the category balance, we propose a stride-based adaptive cropping operation in the preprocessing stage for further boosting the performance of our DeepSRQ method. Next, we present the details of our proposed DeepSRQ. The training and quality prediction steps are also presented.

\subsection{Image Representation}
\begin{figure}[t]
  \centerline{\includegraphics[width=1.0\linewidth]{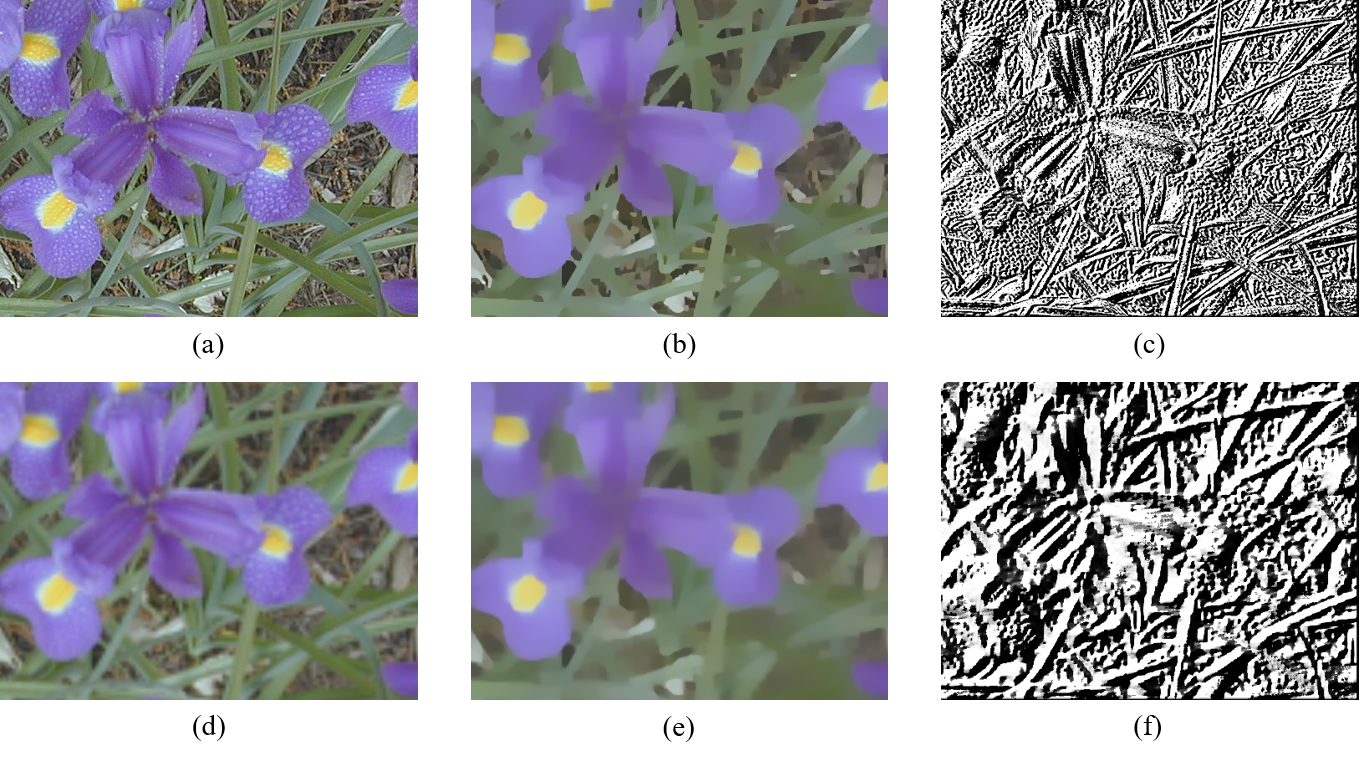}}
  \caption{Examples of natural images and the corresponding structure as well as texture images in the QADS database \cite{zhou2019visual}. (a) Original HR image with higher visual quality. (b) Extracted structure image of (a). (c) Extracted texture image of (a). (d) Distorted SR image with lower visual quality. (e) Extracted structure image of (d). (f) Extracted texture image of (d).}
  \centering
\label{fig:fig3}
\end{figure}

As demonstrated in previous studies, the distortions of SR images cause both image structure and texture degradation \cite{zhou2019visual}; we first extract structure and texture images from distorted SR images. Specifically, we adopt the relative total variation-based structure extraction method \cite{xu2012structure} with default parameters to extract structure images from distorted SR images. From the perspective of perception, the texture map obtained directly by subtracting the structure map from the original image is not necessarily the texture perceived by human eyes. The texture map obtained by the texture descriptors is more consistent with human perception. For texture description, the local binary pattern (LBP) is an effective texture description operator and has significant advantages of rotation invariance and gray invariance \cite{ojala2002multiresolution}. Moreover, the LBP is widely used in image quality evaluation tasks \cite{zhou2017local}. Therefore, we utilize it to extract texture images from distorted SR images. Some examples of natural images and the corresponding structure as well as texture images in the QADS database \cite{zhou2019visual} can be seen in Figure \ref{fig:fig3}. We observe that the structure and texture images can discriminate images with different visual qualities, which demonstrates the effectiveness of the extracted structure and texture images.

To train the two-stream deep neural network, a large quantity of training data is needed. Moreover, we need to use the same fixed sizes during training and testing because fully connected layers exist in the proposed network, and their number of parameters is not flexible. Since the existing publicly available image superresolution quality databases are small-scale, cropping SR images is an effective method for increasing the quantity of training data. Moreover, compared with resizing, cropping ensures that the perceptual image quality is unchanged \cite{kim2017deep}. Hence, we choose to crop multiple patches from different spatial locations to cover the local visual information of the whole SR image without introducing any geometric deformation. The resolution of SR images in the SR quality database \cite{ma2017learning} and QADS database \cite{zhou2019visual} is fixed for different scaling factors and the corresponding Gaussian kernel widths. Therefore, the total number of nonoverlapping cropped patches for each SR image is given by:
\begin{equation}\label{1}
Nu{{m}_{p}}=\left\lfloor \frac{M}{m} \right\rfloor \times \left\lfloor \frac{N}{n} \right\rfloor,
\end{equation}
where $M$ and $m$ are the SR image width and cropped patch width, respectively, while $N$ and $n$ are the SR image height and cropped patch height, respectively. In other words, $M\times N$ is the SR image resolution, and $m\times n$ denotes the cropped patch resolution. Note that $M>m$, $N>n$, and $m=n$ in our experiments.

\subsection{Stride-Based Adaptive Cropping Method}
It should be noted that there exist three amplification factors (i.e., 2, 4, and 8) in the SRID \cite{wang2017perceptual}. A larger amplification factor leads to a higher resolution of SR images. Therefore, we propose a stride-based adaptive cropping approach to ensure the category balance for different amplification factors. Specifically, the stride of the maximum amplification factor (i.e., ${{f}_{\max }}$) is the cropped patch size denoted by $m$. Then, the strides for the other amplification factors (i.e., $f$) are computed as:
\begin{equation}\label{2}
Strid{{e}_{f}}=\frac{f}{{{f}_{\max }}} \times m.
\end{equation}

\begin{figure}[t]
  \centerline{\includegraphics[width=0.6\linewidth]{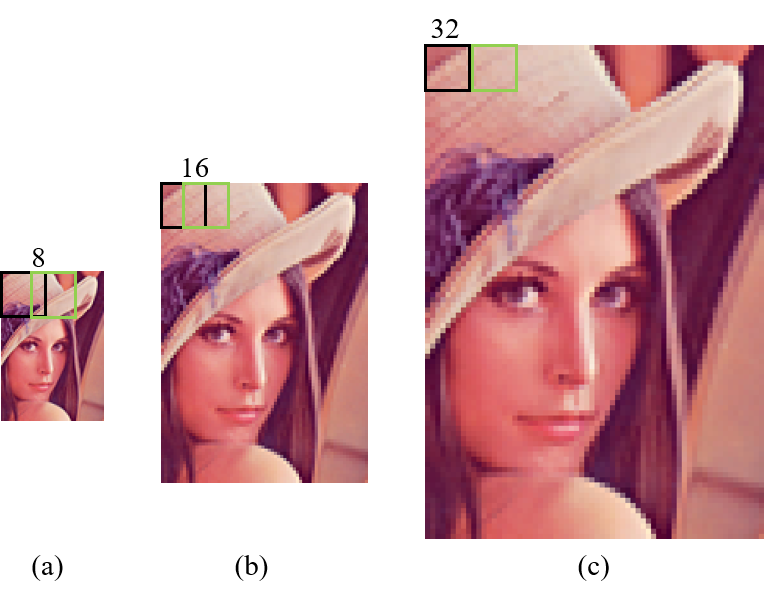}}
  \caption{Illustration of the proposed stride-based adaptive cropping method with several SR image examples, where $8$, $16$, and $32$ denote the stride sizes for SR images with amplification factors $2$, $4$, and $8$, respectively. The cropping patch size is $32\times 32$. (a) SR image with amplification factors equal to $2$. (b) SR image with amplification factors equal to $4$. (c) SR image with amplification factors equal to $8$.}
  \centering
\label{fig:fig4}
\end{figure}

Specifically, as shown in Figure \ref{fig:fig4}, we illustrate several SR image examples with three amplification factors. As the cropping patch size is $32\times 32$, the stride sizes are 8, 16, and 32 for SR images with amplification factors equal to 2, 4, and 8, respectively.
By using the proposed stride-based adaptive cropping approach, the quantity of training data for each amplification factor generally remains balanced. Furthermore, our ablation experiments show that this adaptive cropping approach brings about improved performance.

\subsection{Network Architecture}
Given a distorted SR image $I$, we first extract the structure and texture images from it. Then, we can represent the structure and texture images by a set of cropped patches. Let $\{{{s}_{1}},{{s}_{2}},...,{{s}_{Nu{{m}_{p}}}}\}$ and $\{{{t}_{1}},{{t}_{2}},...,{{t}_{Nu{{m}_{p}}}}\}$ be the cropped structure and texture patches used to train the DCNN. As shown in Figure \ref{fig:fig2}, we first take distorted structure and texture SR image patches as inputs. Following local normalization \cite{ye2012unsupervised}, we rescale each patch to the range [0, 1] by dividing all channels by 255 before feeding it into the two-stream network. The last fully connected layer of each subnetwork is concatenated to obtain a $256-dim$ feature vector, and then two fully connected layers with sizes equal to 256 and 1 are employed to regress the inputs onto one single visual quality score.

\begin{table}[t]
\centering
\caption{Detailed subnetwork configurations of the proposed method. CONV: convolution, ELU: activation, POOL: max pooling, DENSE: fully connected, DROP: dropout.}
\label{table1}
\scalebox{0.8}{
\begin{tabular}{c|c|c}
\hline
Layer & Output Shape & Parameter Number \\ \hline
INPUT & (32, 32, 3) & 0 \\ \hline
CONV1+ELU1 & (32, 32, 16) & 448 \\ \hline
POOL1 & (16, 16, 16) & 0 \\ \hline
CONV2+ELU2 & (16, 16, 16) & 2320 \\ \hline
POOL2 & (8, 8, 16) & 0 \\ \hline
CONV3+ELU3 & (8, 8, 32) & 4640 \\ \hline
CONV4+ELU4 & (8, 8, 32) & 9248 \\ \hline
CONV5+ELU5 & (8, 8, 64) & 18496 \\ \hline
POOL5 & (4, 4, 64) & 0 \\ \hline
DENSE1+ELU6+DROP1 & 128 & 131200 \\ \hline
DENSE2+ELU7+DROP2 & 128 & 16512 \\ \hline
DENSE3 & 1 & 129 \\ \hline
\end{tabular}}
\end{table}

Moreover, we show the ablation substream of our proposed two-stream network. The detailed subnetwork configurations of the proposed method can be found in Table \ref{table1}. Inspired by the work in \cite{krizhevsky2012imagenet}, the designed framework of the subnetwork consists of 12 layers, which include one input layer, five convolutional layers, three max pooling layers, and three fully connected layers. First, for the convolutional layers Conv1 and Conv2, the kernel number is 16. The kernel number is 32 for the convolutional layers Conv3 and Conv4. The kernel number of the last convolutional layer (i.e., Conv5) is 64. Note that each kernel size for the convolutional layers is set to $3\times 3$. We also adopt padding to ensure the unchanged patch size during the process of convolution operations.

In addition, we apply the exponential linear unit (ELU) instead of traditional activation functions such as the sigmoid, tanh, and rectified linear unit (ReLU) after the convolution layer. Formally, the output of the nonlinear activation function ELU can be represented by:
\begin{equation}\label{3}
u(x)=\left\{
\begin{aligned}
& x, x\ge 0 \\
& \alpha({{e}^{x}}-1), x<0 \\
\end{aligned}
\right.,
\end{equation}
where $\alpha $ denotes the parameter to control negative factors and can output information even if the input is negative. Moreover, the mean of the overall output is approximately 0, which is more robust than other traditional activation functions. After the convolution and ELU layers, we exploit max pooling with a window size of $2\times 2$. Therefore, for the feature extractor, the output shape of each feature map is 1/8 of the original input patch size.

We then flatten these feature maps to obtain $128-dim$ feature vectors that can represent the input distorted structure or texture SR image patches. To conduct the perceptual quality regression, we use three fully connected layers with sizes equal to 128, 128, and 1. The ELU layers are also employed after the first two fully connected layers. Meanwhile, dropout is utilized to avoid overfitting. Specifically, the outputs of neurons are set to zero randomly with a particular probability. In our experiments, the probability is either 0.35 or 0.5. The last fully connected layer (i.e., the output layer) has one dimension that represents the predicted quality score.

Finally, to estimate the entire SR image quality, we assume that the visual distortions in reconstructed SR images are roughly homogeneous, which is appropriate for most practical situations. In this case, we thus take estimated scores of image patches as inputs and outputs the average mean of these scores, which is computed as the perceptual quality for the whole SR image.

\subsection{Learning Quality}
We need to obtain a large quantity of training data to train the proposed DeepSRQ network. Meanwhile, the input sizes should be fixed. Hence, we train our network on $32\times 32$ cropped patches extracted from relatively larger distorted structure and texture SR images with the stride-based adaptive cropping approach. The subjective quality scores are taken as ground-truth labels. During the training process, the learning objective function of the network with weights $\textbf{w}$ and updated weights $\widehat{\textbf{w}}$ is defined to minimize the mean squared error (MSE) as follows:
\begin{equation*}
{{\mathcal{L}}_{MSE}}=||{{c}_{\textbf{w}}}({{s}_{i}}, {{t}_{i}})-{{y}_{i}}||_{2}^{2}\
\end{equation*}
\begin{equation}\label{4}
\widehat{\textbf{w}}=\underset{\textbf{w}}{\mathop{\min }}\,L,
\end{equation}
where ${{y}_{i}}$ represents the label of the distorted SR image patch, which is the human rating score. ${{c}_{w}}({{p}_{i}}, {{t}_{i}})$ denotes the quality score computed by the proposed DeepSRQ method. In addition, ${{s}_{i}}\in \{{{s}_{1}},{{s}_{2}},...,{{s}_{Nu{{m}_{p}}}}\}$, ${{t}_{i}}\in \{{{t}_{1}},{{t}_{2}},...,{{t}_{Nu{{m}_{p}}}}\}$, and ${{y}_{i}}\in \{{{y}_{1}},{{y}_{2}},...,{{y}_{Nu{{m}_{p}}}}\}$. ${{Num}_{p}}$ is the total number of input image patches.

To estimate the whole SR image quality by the predicted quality scores of input image patches, we adopt the average quality pooling as:
\begin{equation}\label{5}
Q=\frac{1}{Nu{{m}_{p}}}\sum\limits_{i=1}^{Nu{{m}_{p}}}{{{c}_{\textbf{w}}}({{s}_{i}}, {{t}_{i}})},
\end{equation}
where $Q$ is the final perceptual quality prediction for the SR image. Note that using average pooling for quality evaluation tasks is a common practice because the spatial distortion in SR images is generally homogeneous. Indeed, we also test different saliency models, but the performance does not improve. One possible explanation is that state-of-the-art saliency detection algorithms focus on objects rather than distortion, which is more important for image quality assessment.

\section{Experimental Results and Analysis}
In this section, we report the results of several experiments to test the performance of DeepSRQ on three existing publicly available image superresolution quality databases, i.e., the SR quality database \cite{ma2017learning}, SRID \cite{wang2017perceptual}, and QADS \cite{zhou2019visual}. We also pretrain DeepSRQ on the SR quality database, then perform cross validation on SRID, and vice versa. Moreover, we examine the effects of several parameter settings, including patch size and kernel size, and visualize the feature map to discover what has been learned from our proposed two-stream deep learning architecture. Furthermore, we carry out ablation experiments to test and quantify the performance gain of each key technique for learning the perceptual quality of image superresolution.

\subsection{Protocol}
We briefly introduce three publicly available image superresolution quality databases and the three commonly used criteria employed in the experiments.

The \textbf{SR quality database} includes a total of 1,620 SR images that are generated from LR images by nine SR algorithms. These SR algorithms are applied with a variety of scaling factors and kernel widths, denoted by s and $\sigma $, respectively. The numbers of ground-truth HR images and LR images are 30 and 60, respectively. Note that the larger subsampling factor requires a larger blur kernel width for better performance. Therefore, the optimal kernel width is applied for each scaling factor. The parameter selection details of this database can be found in Table \ref{table2}. A subjective experiment is conducted to collect the subjective quality scores from 50 subjects. The mean of the median 40 subject scores is computed as the ground truth in the form of the mean opinion score (MOS), ranging from 0 to 10. Here, higher MOS means better perceptual quality.

\textbf{SRID} consists of 480 distorted SR images that are directly generated by LR images using two interpolation methods and six SR enhancement algorithms with three amplification factors of 2, 4, and 8. Nondistorted HR images are unavailable in this database. Subjective quality scores are provided in the form of MOS ranging from 0 to 10; the higher the value, the better the perceptual quality.

\textbf{QADS} contains 20 source images and 980 SRIs. The source images are selected from the MDID database \cite{sun2017mdid} and Set14 database \cite{zeyde2010single}. Three magnification scales are introduced to obtain the 980 SRIs, including 2 times, 3 times, and 4 times. Twenty-one image superresolution algorithms are applied to obtain the distorted SR images. Subjective quality ratings are given in MOS ranging from 0 to 1. A higher value indicates better perceptual quality.

\begin{table}[t]
\centering
\caption{The values of scaling factors ($s$) and the corresponding kernel width values ($\sigma $) in SR quality database \cite{ma2017learning}.}
\label{table2}
\scalebox{0.8}{
\begin{tabular}{c|c|c|c|c|c|c}
\hline
\textbf{$s$}       & 2 & 3 & 4 & 5 & 6 & 8 \\ \hline
\textbf{$\sigma $} & 0.8 & 1.0 & 1.2 & 1.6 & 1.8 & 2.0 \\ \hline
\end{tabular}}
\end{table}

We adopt three commonly used criteria to evaluate the performance of DeepSRQ: Spearman rank-order correlation coefficient (SROCC), Pearson linear correlation coefficient (PLCC), and root mean square error (RMSE). Here, SROCC is used to evaluate prediction monotonicity, while PLCC and RMSE are used to evaluate prediction accuracy. Higher correlation coefficients and lower error indicate better agreement with human quality ratings. Moreover, before calculating the PLCC and RMSE performance of objective quality assessment methods, a nonlinear logistic fitting is applied to map the predicted scores to the same scales of subjective quality scores. Following \cite{video2003final}, we adopt a four-parameter logistic function as follows:
\begin{equation}\label{7}
g(x)=\frac{{{\tau }_{1}}-{{\tau }_{2}}}{1+{{e}^{\frac{x-{{\tau }_{3}}}{{{\tau }_{4}}}}}}+{{\tau }_{2}},
\end{equation}
where ${{\tau }_{1}}$ to ${{\tau }_{4}}$ are four free parameters to be determined in the curve fitting process. $x$ denotes the raw objective score, and $g(x)$ represents the mapped score after the fitting.

\subsection{Training Details}
In the experiments, we use the stochastic gradient descent (SGD) as the optimization algorithm with 0.9 momentum. The learning rate is initially set to ${{10}^{\text{-}2}}$ with ${{10}^{\text{-}6}}$ decay. We update the network weights through backpropagation. The batch size is 128 in our experiments. For each image superresolution quality database, we randomly select 80\% image data as the training set and the remaining 20\% for testing. There is no overlap of source image content between the training and testing sets. The performance of the proposed DeepSRQ method is reported after 1,000 epochs.

\subsection{Baselines}
The proposed DeepSRQ method not only combines feature extraction with quality regression as well as pooling evaluation in a joint learning process but also creates synthetically learned features (i.e., both low-level visual information and high-level semantic features). To verify the effectiveness of our proposed method, it is compared with state-of-the-art IQA metrics using the handcrafted low-level features and the high-level semantic features extracted from pretrained DCNN models.

Due to the existence of original distortion-free HR images in the SR quality database \cite{ma2017learning}, we compare our method with both the FR-IQA (i.e., PSNR, SSIM \cite{wang2004image}, multiscale SSIM denoted by MS-SSIM \cite{wang2003multiscale}) and NR-IQA algorithms, which include the blind/referenceless image spatial quality evaluator (BRISQUE) \cite{mittal2012no}, natural image quality evaluator (NIQE) \cite{mittal2013making}, local natural image quality evaluator (ILNIQE) \cite{zhang2015feature}, convolutional neural networks for no-reference image quality assessment (CNN-IQA) \cite{kang2014convolutional}, and the shallow convolutional neural network for SR IQA (CNNSR) \cite{fang2018blind}. Moreover, we employ the pretrained DCNN (i.e., the well-known ResNet50 \cite{he2016deep}) to extract high-level semantic features using the Caffe framework \cite{jia2014caffe}, and then combine the high-level semantic features (i.e., the 2,048 dimensions output of the pool5 layer) with other handcrafted features such as the NSS features. We then input these features into the support vector regression (SVR) model to predict the quality score.

For the QADS database \cite{zhou2019visual}, we compare the proposed DeepSRQ with 15 types of both traditional FR-IQA and NR-IQA metrics, including PSNR, SSIM \cite{wang2004image}, MS-SSIM \cite{wang2003multiscale}, information fidelity criterion (IFC) \cite{sheikh2005information}, visual information fidelity (VIF) \cite{sheikh2004image}, most apparent distortion (MAD) \cite{larson2010most}, information content weighted SSIM (IW-SSIM) \cite{wang2011information}, feature similarity (FSIM) index \cite{zhang2011fsim}, gradient similarity (GSIM) index \cite{liu2012image}, internal generative mechanism (IGM) \cite{wu2013perceptual}, gradient magnitude similarity deviation (GMSD) \cite{xue2014gradient}, directional anisotropy structure measurement (DASM) \cite{ding2017image}, superpixel-based similarity (SPSIM) index \cite{sun2018spsim}, structure-texture decomposition-based IQA approach called SIS \cite{zhou2019visual}, and local pattern statistics index (LPSI) \cite{wu2015highly}. It should be noted that the SIS \cite{zhou2019visual} is an FR method that considers structure and texture information based on traditional handcrafted features. However, our proposed DeepSRQ is a DCNN-based NR algorithm. Therefore, we also compare the proposed DeepSRQ with two deep learning-based image quality evaluation methods, namely, CNN-IQA \cite{kang2014convolutional} and deep bilinear CNN (DBCNN) \cite{zhang2018blind}.

Since the SRID database \cite{wang2017perceptual} applies three different amplification factors to generate the SR images, the resolutions of these SR images vary. Moreover, the input size of ResNet50 is fixed to $224\times 224$. Therefore, only the performance result of this method on the SR quality database \cite{ma2017learning} is shown. In addition, note that the SRID database has no-reference HR images; thus, more NR-IQA metrics are used for performance comparison except for those used on the SR quality database, which include no-reference free energy-based robust metric (NFERM) \cite{gu2015using}, blind image quality index (BIQI) \cite{moorthy2010two}, blind image integrity notator using DCT statistics (BLIINDS-II) \cite{saad2012blind}, codebook representation for no-reference image assessment (CORNIA) \cite{ye2012unsupervised}, derivative statistics-based image quality evaluator (DESIQUE) \cite{zhang2013no}, distortion identification-based image verity and integrity evaluation index (DIIVINE) \cite{moorthy2011blind}, and six-step blind metric (SISBLIM) \cite{gu2014hybrid}.

In addition, we compare the proposed DeepSRQ with several state-of-the-art image sharpness assessment methods, including spectral and spatial sharpness measure (S3) \cite{vu2011bf}, local phase coherence-based sharpness index (LPC-SI) \cite{hassen2013image}, HVS-MaxPol-1 \cite{hosseini2019encoding} using the best single kernel, and HVS-MaxPol-2 \cite{hosseini2019encoding} adopting the combination of the best two kernels.

\subsection{Performance Comparison}
In this part, three publicly available SR image quality databases \cite{ma2017learning,wang2017perceptual} are used for performance comparison. For the SR quality database \cite{ma2017learning}, we compare the performance of our proposed DeepSRQ method with three classical FR-IQA metrics, namely PSNR, SSIM \cite{wang2004image}, and MS-SSIM \cite{wang2003multiscale}. Additionally, several state-of-the-art NR-IQA metrics, including BRISQUE \cite{mittal2012no}, NIQE \cite{mittal2013making}, ILNIQE \cite{zhang2015feature}, CNN-IQA \cite{kang2014convolutional} and CNNSR \cite{fang2018blind}, are taken for performance comparison. Among these four NR-IQA metrics, the CNNSR \cite{fang2018blind} is a shallow convolutional neural network specifically designed for evaluating the quality of SR images. As shown in Table \ref{table3}, our proposed DeepSRQ outperforms both the FR and NR algorithms. The reason why the RMSE values of BRISQUE and ILNIQUE are much larger might be that they are not designed, optimized and tested for image superresolution applications.

\begin{table}[t]
\centering
\caption{Performance comparison on SR quality database \cite{ma2017learning}. FR: full-reference, NR: no-reference.}
\label{table3}
\scalebox{0.8}{
\begin{tabular}{c|c|c|c|c}
\hline
Type & Method & SROCC & PLCC & RMSE \\
\hline
\multirow{3}{*}{FR}    & PSNR & 0.3110 & 0.3335 & 2.9383 \\
                       & SSIM \cite{wang2004image} & 0.5562 & 0.5726 & 1.7980 \\
                       & MS-SSIM \cite{wang2003multiscale} & 0.6452 & 0.6218 & 1.0272 \\
\hline
\multirow{10}{*}{NR}   & BRISQUE \cite{mittal2012no} & 0.5721 & 0.6176 & 10.0747 \\
                       & NIQE \cite{mittal2013making} & 0.6254 & 0.6364 & 1.5582 \\
                       & ILNIQE \cite{zhang2015feature} & 0.6282 & 0.6198 & 18.3748 \\
                       & LPSI \cite{wu2015highly} & 0.4896 & 0.5276 & 2.0422 \\
                       & S3 \cite{vu2011bf} & 0.5066 & 0.5494 & 2.0087 \\
                       & LPC-SI \cite{hassen2013image} & 0.5441 & 0.5665 & 1.9812 \\
                       & HVS-MaxPol-1 \cite{hosseini2019encoding} & 0.6423 & 0.6706 & 1.7834 \\
                       & HVS-MaxPol-2 \cite{hosseini2019encoding} & 0.6314 & 0.6417 & 1.8438 \\
                       & CNN-IQA \cite{kang2014convolutional} & 0.7983 & 0.8398 & 1.312 \\
                       & CNNSR \cite{fang2018blind} & 0.8394 & 0.9156 & 1.2527 \\
                       & ResNet50-pool5+NSS+SVR  & 0.8734 & 0.8873 & 1.1060 \\
                       & \textbf{Proposed DeepSRQ} & \textbf{0.9206} & \textbf{0.9273} & \textbf{0.9042} \\
\hline
\end{tabular}}
\end{table}

\begin{table}[t]
\centering
\caption{Performance comparison on QADS database \cite{zhou2019visual}. FR: full-reference, NR: no-reference.}
\label{table4}
\scalebox{0.8}{
\begin{tabular}{c|c|c|c|c}
\hline
Type & Method & SROCC & PLCC & RMSE \\
\hline
\multirow{14}{*}{FR}   & PSNR & 0.3544 & 0.3897 & 0.2530 \\
                       & SSIM \cite{wang2004image} & 0.5290 & 0.5327 & 0.2325 \\
                       & MS-SSIM \cite{wang2003multiscale} & 0.7172 & 0.7240 & 0.1895 \\
                       & IFC \cite{sheikh2005information} & 0.8609 & 0.8657 & 0.1375 \\
                       & VIF \cite{sheikh2004image} & 0.8152 & 0.8210 & 0.1568 \\
                       & MAD \cite{larson2010most} & 0.7234 & 0.7311 & 0.1874 \\
                       & IW-SSIM \cite{wang2011information} & 0.8195 & 0.8234 & 0.1559 \\
                       & FSIM \cite{zhang2011fsim}  & 0.6885 & 0.6902 & 0.1988 \\
                       & GSIM \cite{liu2012image} & 0.5538 & 0.5684 & 0.2260 \\
                       & IGM \cite{wu2013perceptual} & 0.7145 & 0.7192 & 0.1907 \\
                       & GMSD \cite{xue2014gradient} & 0.7650 & 0.7749 & 0.1736 \\
                       & DASM \cite{ding2017image} & 0.7512 & 0.7585 & 0.1790 \\
                       & SPSIM \cite{sun2018spsim} & 0.5751 & 0.5822 & 0.2233 \\
                       & SIS \cite{zhou2019visual} & 0.9232 & 0.9230 & 0.1057 \\
\hline
\multirow{5}{*}{NR}    & LPSI \cite{wu2015highly} & 0.4051 & 0.4207 & 0.2492 \\
                       & S3 \cite{vu2011bf} & 0.4636 & 0.4671 & 0.2429 \\
                       & LPC-SI \cite{hassen2013image} & 0.4902 & 0.4846 & 0.2403 \\
                       & HVS-MaxPol-1 \cite{hosseini2019encoding} & 0.6160 & 0.6169 & 0.2162 \\
                       & HVS-MaxPol-2 \cite{hosseini2019encoding} & 0.5739 & 0.5817 & 0.2234 \\
                       & CNN-IQA \cite{kang2014convolutional} & 0.8665 & 0.8709 & 0.1280 \\
                       & DBCNN \cite{zhang2018blind} & 0.8707 & 0.8589 & 0.1508 \\
                       & \textbf{Proposed DeepSRQ} & \textbf{0.9528} & \textbf{0.9557} & \textbf{0.0767} \\
\hline
\end{tabular}}
\end{table}

\begin{table}[t]
\centering
\caption{Performance comparison on the SRID database \cite{wang2017perceptual}.}
\label{table5}
\scalebox{0.8}{
\begin{tabular}{c|c|c|c}
\hline
Method & SROCC & PLCC & RMSE \\ \hline
BRISQUE\cite{mittal2012no} & 0.6666 & 0.6738 & 1.1953 \\ \hline
NIQE \cite{mittal2013making} & 0.4759 & 0.5247 & 1.3769 \\ \hline
ILNIQE \cite{zhang2015feature} & 0.4233 & 0.4136 & 1.4729 \\ \hline
NFERM \cite{gu2015using} & 0.6177 & 0.6011 & 1.2927 \\ \hline
BIQI \cite{moorthy2010two} & 0.4336 & 0.4253 & 1.2682 \\ \hline
BLIINDS-II \cite{saad2012blind} & 0.3687 & 0.3783 & 1.4973 \\ \hline
CORNIA \cite{ye2012unsupervised} & 0.5985 & 0.6767 & 1.1909 \\ \hline
DESIQUE \cite{zhang2013no} & 0.5453 & 0.5253 & 1.3763 \\ \hline
DIIVINE \cite{moorthy2011blind} & 0.4826 & 0.4286 & 1.4614 \\ \hline
SISBLIM \cite{gu2014hybrid} & 0.5965 & 0.6223 & 1.2661 \\ \hline
LPSI \cite{wu2015highly} & 0.7454 & 0.7457 & 1.0777 \\ \hline
S3 \cite{vu2011bf} & 0.1797 & 0.1800 & 1.5910 \\ \hline
LPC-SI \cite{hassen2013image} & 0.0234 & 0.1978 & 1.6613 \\ \hline
HVS-MaxPol-1 \cite{hosseini2019encoding} & 0.3736 & 0.3307 & 1.5264 \\ \hline
HVS-MaxPol-2 \cite{hosseini2019encoding} & 0.4561 & 0.4237 & 1.4651 \\ \hline
CNN-IQA \cite{kang2014convolutional} & 0.8541 & 0.8783 & 0.7753 \\ \hline
DBCNN \cite{zhang2018blind} & 0.6439 & 0.7422 & 4.5729 \\ \hline
\textbf{Proposed DeepSRQ} & \textbf{0.9138} & \textbf{0.9309} & \textbf{0.5922} \\ \hline
\end{tabular}}
\end{table}

Furthermore, since DeepSRQ considers both low-level visual information and high-level semantic features, we experimentally show that the synthetically learned features are more effective than both the handcrafted low-level features and the high-level semantic features extracted from pretrained DCNN models. Specifically, we employ the remarkable residual learning-based network (i.e., ResNet50) to extract high-level semantic features from its pool5 layer. In addition, before extracting the features, we crop the SR images into patches with $224\times 224$ pixels due to the fixed input size of ResNet50. For each cropped image patch, we then obtain a $2,048-dim$ feature. Since each patch in a particular SR image is equally important to the contribution of final perceptual quality, we input the average values of these high-level semantic features and the handcrafted low-level features (i.e., NSS features) into the SVR model for predicting the quality scores of SR images. The database is also divided randomly into 80\% for training and 20\% for testing. Finally, the procedure is repeated 1,000 times, and the median values are taken as the experimental results reported in Table \ref{table3}. We find that our proposed DeepSRQ outperforms ResNet50-pool5+NSS+SVR, which further demonstrates the effectiveness of synthetically learned features in a deep neural network.

For the QADS \cite{zhou2019visual}, we compare the proposed DeepSRQ with state-of-the-art FR-IQA and NR-IQA metrics. The performance comparison results are shown in Table \ref{table4}, which demonstrate that our proposed DeepSRQ method outperforms the other FR-IQA approaches. The compared FR-IQA algorithms include PSNR, SSIM \cite{wang2004image}, MS-SSIM \cite{wang2003multiscale}, IFC \cite{sheikh2005information}, VIF \cite{sheikh2004image}, MAD \cite{larson2010most}, IW-SSIM \cite{wang2011information}, FSIM \cite{zhang2011fsim}, GSIM \cite{liu2012image}, IGM \cite{wu2013perceptual}, GMSD \cite{xue2014gradient}, DASM \cite{ding2017image}, SPSIM \cite{sun2018spsim}, SIS \cite{zhou2019visual}, and LPSI \cite{wu2015highly}. Note that the SIS adopts a structure-texture decomposition method and then calculates similarities from textural, structural and high-frequency aspects to form a parametric model. Our proposed DeepSRQ performs better than this SIS method due to the powerful learned discriminative features from the two-stream network. Moreover, the proposed DeepSRQ outperforms state-of-the-art NR-IQA methods, such as CNN-IQA \cite{kang2014convolutional} and DBCNN \cite{zhang2018blind}. This is mainly because the characteristics of SR images are not well considered in these algorithms.

For the SRID \cite{wang2017perceptual}, since the originally nondistorted HR images are unavailable in this database, we compare our method with more state-of-the-art NR-IQA metrics, which include BRISQUE \cite{mittal2012no}, NIQE \cite{mittal2013making}, ILNIQE \cite{zhang2015feature}, NFERM \cite{gu2015using}, BIQI \cite{moorthy2010two}, BLIINDS-II \cite{saad2012blind}, CORNIA \cite{ye2012unsupervised}, DESIQUE \cite{zhang2013no}, DIIVINE \cite{moorthy2011blind}, and SISBLIM \cite{gu2014hybrid}. The performance comparison values are provided in Table \ref{table5}. It can be seen that the proposed DeepSRQ method outperforms the other NR-IQA metrics.

In all three adopted SR image quality databases, several state-of-the-art image sharpness assessment methods are compared with our proposed DeepSRQ. As shown in Tables \ref{table3}, \ref{table4} and \ref{table5}, the proposed DeepSRQ can achieve better performance on all SR image quality databases.

\subsection{Cross Dataset Validation}
In addition, we test the generalization ability of our proposed DeepSRQ method through cross dataset validation. Since a similar data distribution is assumed between the training and testing images, we report the performance on the three image superresolution quality databases \cite{ma2017learning,wang2017perceptual}.

\begin{table}[t]
\centering
\caption{Performance values in cross dataset evaluation.}
\label{table6}
\scalebox{0.8}{
\begin{tabular}{c|c|c|c}
\hline
Train $\rightarrow$ Test  & SROCC & PLCC & RMSE \\ \hline
SR quality database $\rightarrow$ SRID & 0.7225 & 0.7486 & 1.0749 \\ \hline
SRID $\rightarrow$ SR quality database & 0.8431 & 0.8415 & 1.3055 \\ \hline
\end{tabular}}
\end{table}

\begin{figure}[t]
  \centerline{\includegraphics[width=0.6\linewidth]{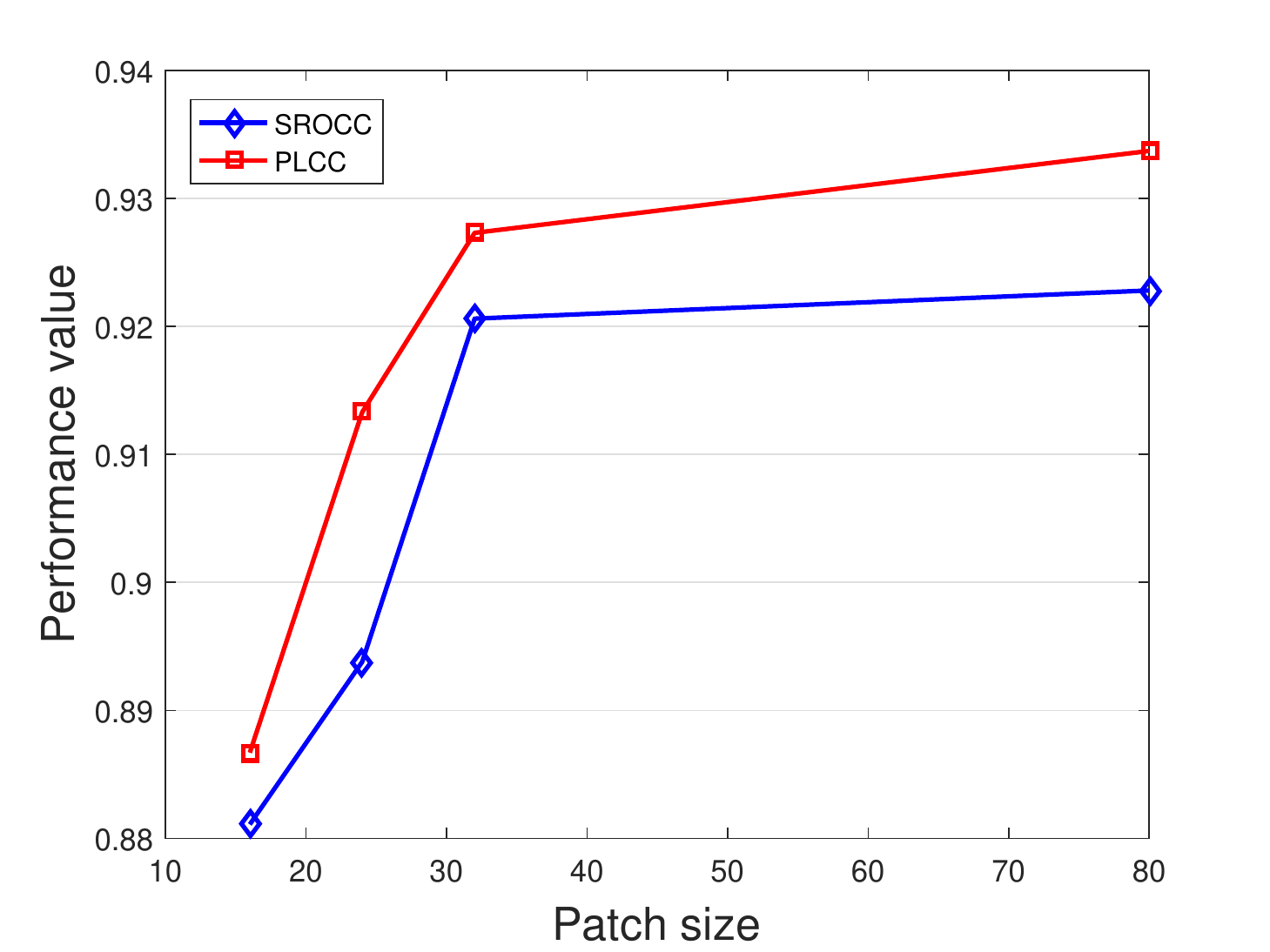}}
  \caption{SROCC and PLCC performance with respect to the patch sizes including $16\times 16$, $24\times 24$, $32\times 32$, and $80\times 80$ on SR quality database \cite{ma2017learning}.}
  \centering
\label{fig:fig5}
\end{figure}

\begin{figure}[t]
  \subfigure[]{ \label{figure:a}
  \includegraphics[width=0.5\linewidth]{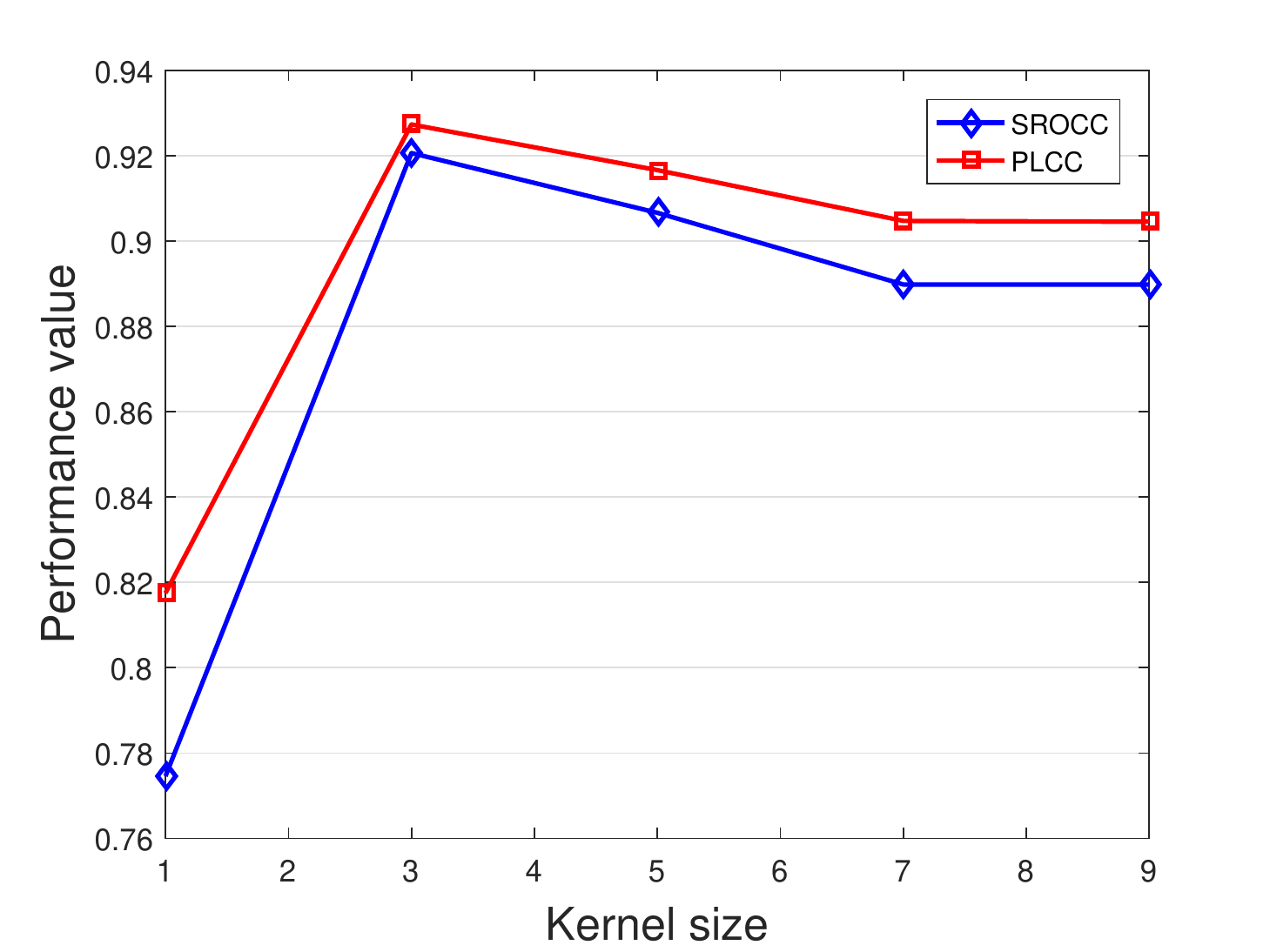}
  }
  \subfigure[]{ \label{figure:b}
  \includegraphics[width=0.5\linewidth]{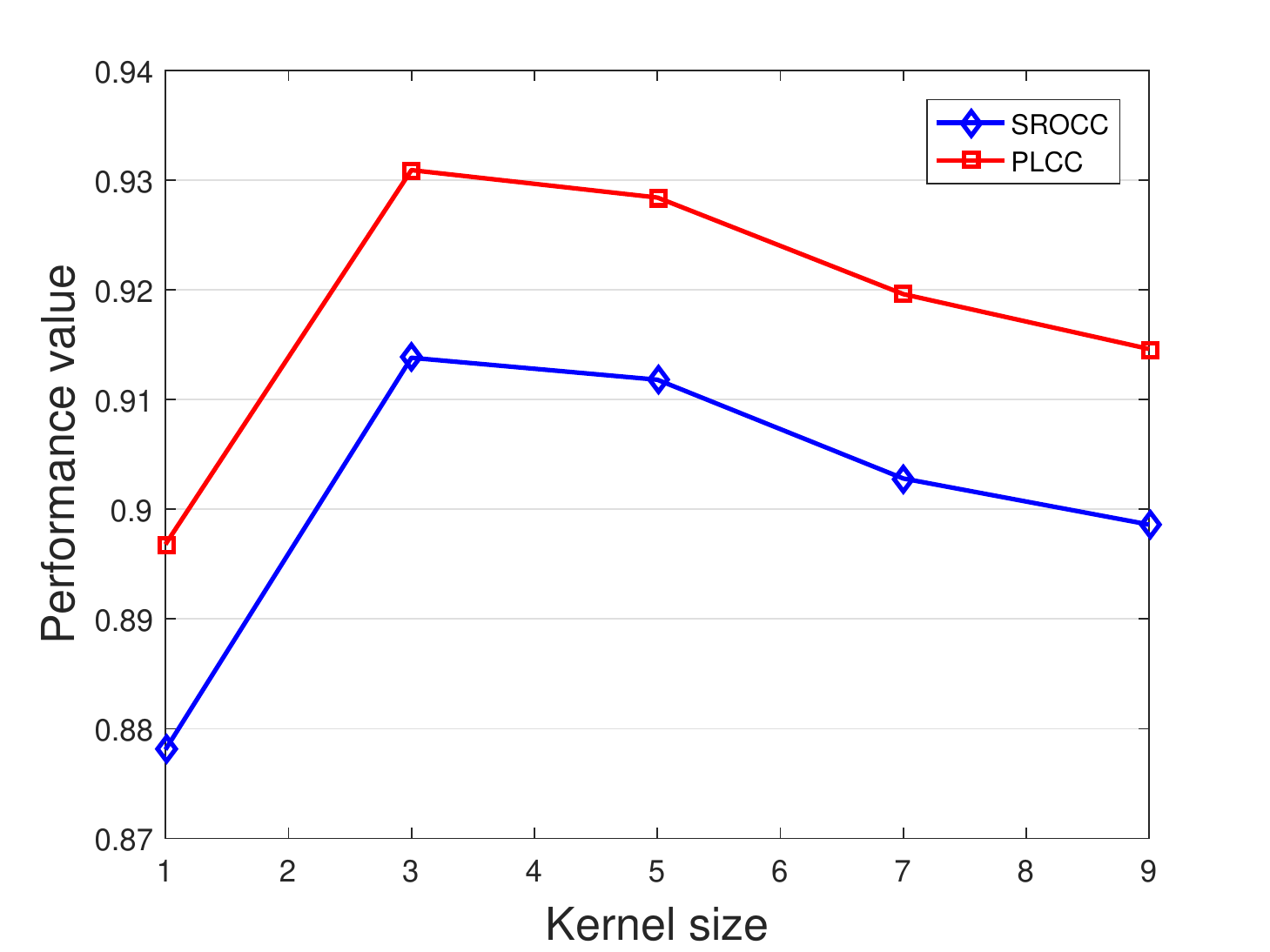}
  }
  \caption{SROCC and PLCC performance with respect to the kernel sizes ranging from $1\times 1$ to $9\times 9$. (a) Run on SR quality database \cite{ma2017learning}. (b) Run on SRID \cite{wang2017perceptual}.}
  \centering
  \label{fig:fig67}
\end{figure}

\begin{figure}[t]
  \centerline{\includegraphics[width=0.6\linewidth]{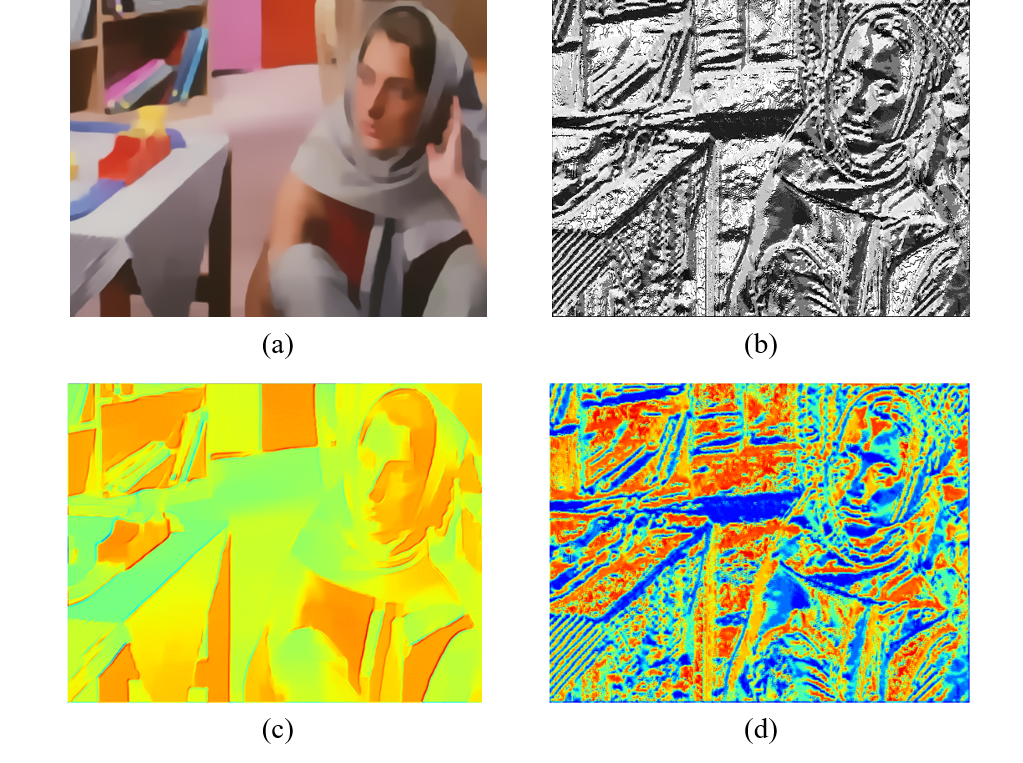}}
  \caption{Examples of distorted structure and texture SR images as well as the corresponding feature maps in the first convolutional layer. (a) Distorted structure SR image. (b) Distorted texture SR image. (c) Feature map of (a). (d) Feature map of (b).}
  \centering
\label{fig:fig9}
\end{figure}

As shown in Table \ref{table6}, DeepSRQ has a promising generalization ability for different databases. In other words, our proposed DeepSRQ is independent and robust for the used image superresolution quality databases. Note that the reason for the performance difference is due to different distributions in the three SR image quality databases.

\subsection{Effects of Parameters}
Several network parameters are involved in the proposed DeepSRQ design. To understand how these network parameters affect the performance of our proposed DeepSRQ method, we carry out experiments to test the DeepSRQ with different parameter settings.

Since the cropping approach is applied to increase our training data and the local visual information of the whole SR images is considered in our experiments, we examine how the patch size affects the performance of the proposed DeepSRQ method. Note that a fixed sampling stride (i.e., 32) is used for the SR quality database \cite{ma2017learning} to ensure that the amount of training data remains unchanged. Then, we vary the patch size while fixing the rest of the network architecture to plot the performance for the SR quality database \cite{ma2017learning}.

Figure \ref{fig:fig5} shows the change in performance with respect to the patch sizes including $16\times 16$, $24\times 24$, $32\times 32$, and $80\times 80$. From Figure \ref{fig:fig5}, we can see that a larger patch size results in better performance of the trained network. Moreover, the performance increases slightly as the patch size increases from $32\times 32$ to $80\times 80$. However, a larger patch size not only reduces the spatial quality resolution but also causes more processing time for training. Therefore, we prefer a relatively small patch size that can also yield promising performance.

\begin{table}[t]
\centering
\caption{Ablation study of each substream on SR quality database \cite{ma2017learning}, SRID database \cite{wang2017perceptual}, and QADS database \cite{zhou2019visual}.}
\label{table7}
\scalebox{0.8}{
\begin{tabular}{c|c|c|c}
\hline
SR quality database \cite{ma2017learning} & SROCC & PLCC & RMSE \\ \hline
Structure & 0.8242 & 0.8213 & 1.3787 \\ \hline
Texture & 0.9049 & 0.9153 & 0.9733 \\ \hline
\textbf{Proposed DeepSRQ} & \textbf{0.9206} & \textbf{0.9273} & \textbf{0.9042} \\ \hline
SRID database \cite{wang2017perceptual} & SROCC & PLCC & RMSE \\ \hline
Structure & 0.8619 & 0.8797 & 0.7709 \\ \hline
Texture & 0.8840 & 0.9094 & 0.6742 \\ \hline
\textbf{Proposed DeepSRQ} & \textbf{0.9138} & \textbf{0.9309} & \textbf{0.5922} \\ \hline
QADS database \cite{zhou2019visual} & SROCC & PLCC & RMSE \\ \hline
Structure & 0.9137 & 0.9214 & 0.1012 \\ \hline
Texture & 0.9138 & 0.9242 & 0.0995 \\ \hline
\textbf{Proposed DeepSRQ} & \textbf{0.9528} & \textbf{0.9557} & \textbf{0.0767} \\ \hline
\end{tabular}}
\end{table}

In addition, different kernel sizes in the convolutional layers may lead to various performances because of the receptive field. Therefore, to discover how the kernel size affects the performance of our DeepSRQ algorithm, we change the kernel size while fixing the rest of the network architecture to plot the performance for the SR quality database \cite{ma2017learning} and SRID \cite{wang2017perceptual}.

Figure \ref{fig:fig67} shows the change in SROCC and PLCC performance with respect to the kernel sizes ranging from $1\times 1$ to $9\times 9$. Except for the kernel size of $1\times 1$, we can observe that a small kernel size creates an increase in both SROCC and PLCC performance. One possible explanation is that the relatively small receptive field (i.e., $3\times 3$) is important for the image SR problem, which can effectively capture the notion of five orientations: up, down, left, right, and center.

\subsection{Visualize Learned Feature Map}
To discover what has been learned from the proposed two-stream deep learning scheme, we visualize the feature maps in the first convolutional layer. Figure \ref{fig:fig9} depicts one of the feature maps at the first convolutional layer for both distorted structure and texture SR images. We can observe that the structural information and textural details can be separately learned from the structure and texture substreams, respectively, which further verifies the effectiveness of our DeepSRQ method.

\subsection{Ablation Experiments}
\begin{figure}[t]
  \centerline{\includegraphics[width=1.0\linewidth]{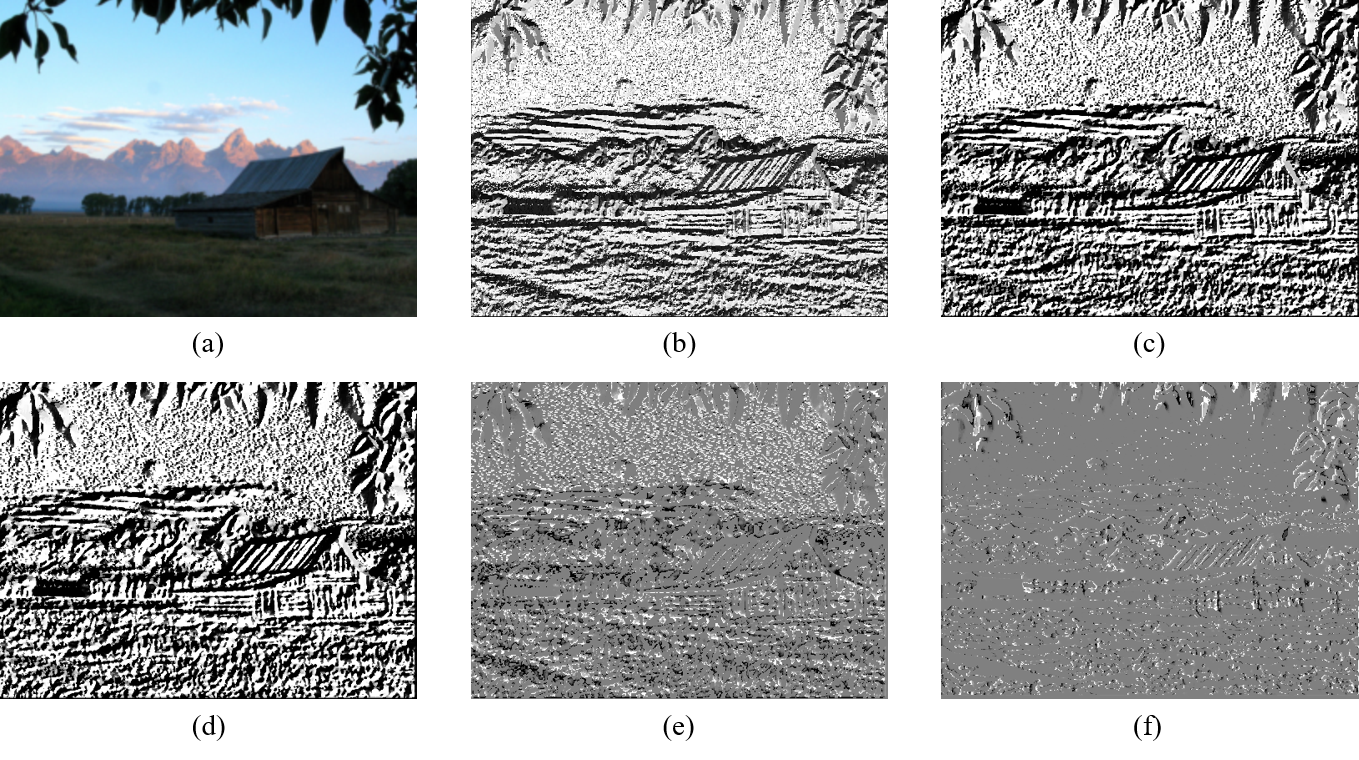}}
  \caption{Examples of distorted SR images and the corresponding texture images with different LBP radii $r$ in the QADS database \cite{zhou2019visual}. (a) Distorted SR image. (b) Extracted texture image with LBP radius $r=1$. (c) Extracted texture image with LBP radius $r=2$. (d) Extracted texture image with LBP radius $r=3$. (e) Extracted texture image with LBP radius $r=4$. (f) Extracted texture image with LBP radius $r=5$.}
  \centering
\label{fig:fig8}
\end{figure}

To validate the necessity of each subnetwork and how they contribute to the whole two-stream framework, we use each substream to perform the perceptual quality prediction of SR images. The results of this ablation study are provided in Table \ref{table7}. We find that the combination of structure and texture streams helps improve the final performance. Moreover, since the LBP texture descriptor involves the radius parameter, which is denoted by $r$, we also vary different values of $r$ to discover how it influences the performance of our algorithm. Table \ref{table8} shows the change in SROCC, PLCC and RMSE performance with respect to the radius $r$ on the QADS database \cite{zhou2019visual}. Examples of distorted SR images and the corresponding texture images with different LBP radii $r$ ranging from 1 to 5 in the QADS database \cite{zhou2019visual} are illustrated in Figure \ref{fig:fig8}. We can observe that a smaller radius brings about an increase in performance due to more reserved texture details.

\begin{table}[t]
\centering
\caption{Parameter experiment results on QADS database \cite{zhou2019visual}.}
\label{table8}
\scalebox{0.8}{
\begin{tabular}{c|c|c|c}
\hline
Parameter Settings & SROCC & PLCC & RMSE \\ \hline
$r=1$ & \textbf{0.9138} & \textbf{0.9242} & \textbf{0.0995} \\ \hline
$r=2$ & 0.9116 & 0.9197 & 0.1023 \\ \hline
$r=3$ & 0.8988 & 0.9028 & 0.1121 \\ \hline
$r=4$ & 0.8809 & 0.8839 & 0.1218 \\ \hline
$r=5$ & 0.8807 & 0.8808 & 0.1233 \\ \hline
\end{tabular}}
\end{table}

\begin{table}[t]
\centering
\caption{Ablation experiment about adaptive cropping approach on SRID database \cite{wang2017perceptual}. DeepSRQ$\backslash$ADA denotes the removed adaptive cropping approach. The best results are in bold.}
\label{table9}
\scalebox{0.8}{
\begin{tabular}{c|c|c|c}
\hline
Ablation & SROCC & PLCC & RMSE \\ \hline
DeepSRQ$\backslash$ADA & 0.8988 & 0.9061 & 0.6736 \\ \hline
\textbf{Proposed DeepSRQ} & \textbf{0.9138} & \textbf{0.9309} & \textbf{0.5922} \\ \hline
\end{tabular}}
\end{table}

Additionally, to demonstrate that the adopted techniques are critical for the performance of DeepSRQ for perception-driven image superresolution, we further conduct several ablation experiments. Specifically, we remove the adaptive cropping approach and then test the performance of the remaining framework. As shown in Table \ref{table9}, the proposed stride-based adaptive cropping approach is validated to further improve the performance of our proposed DeepSRQ.

\section{Conclusions}
In this paper, we propose a two-stream network to predict the perceptual quality of SR images in a no-reference manner, which is demonstrated to be more consistent with human perception. We consider both the structural and textural characteristics of the distortions in SR images. The top performance and promising generalization capacity of our proposed DeepSRQ method are validated by comparison with state-of-the-art IQA algorithms on three publicly available SR image quality databases. Experimental results also show that the synthetically learned features in a deep neural network are more effective than both the handcrafted low-level visual features and the high-level semantic features. Moreover, we validate that the two-stream scheme performs better than each substream through an ablation study. Extensive parameter experiments also show various aspects of our proposed DeepSRQ. In addition, the proposed stride-based adaptive cropping approach is verified to further improve the performance of the proposed DeepSRQ method.

In future studies, we will apply the DeepSRQ metric to automatically optimize image SR frameworks, including both learning-free and learning-based SR algorithms. Furthermore, it is worth designing more effective and robust deep neural networks by considering more relevant characteristics of image superresolution.

\section*{Acknowledgment}
This work was supported in part by NSFC under Grant 61571413, 61632001.

\small
\bibliography{references}

\end{document}